\documentclass{revtex4}

\begin{document}

\title{Dynamics of Inflationary Universes with Positive Spatial Curvature}
\author{G. F. R. Ellis$^{1,2}$, W. Stoeger$^3$, P. McEwan$^1$ 
and P. Dunsby$^1$}
\address{$1$ Department of Applied Mathematics, University of Cape Town, 
Rondebosch 7700, Cape Town, South Africa.}
\address{$2$ Erwin Schroedinger Institute, Vienna, Austria.}
\address{$3$ Vatican Observatory, Tucson, Arizona, USA.}

\date{\today}

\begin{abstract}
If the spatial curvature of the universe is positive, then the curvature
term will always dominate at early enough times in a slow-rolling
inflationary epoch. This enhances inflationary effects and hence puts limits
on the possible number of e-foldings that can have occurred, independently
of what happened before inflation began and in particular without regard for
what may have happened in the Planck era. We use a simple multi-stage model to examine this limit as a function of the present density parameter $\Omega _0$ and the epoch when inflation ends.
\end{abstract}

\maketitle

\section{Positively-Curved Inflationary Models?}

The inflationary universe paradigm \cite{Ref1} is the premiere causative
concept in present-day physical cosmology, and faith in this view has been
bolstered by the recent measurements of a second and third peak in the
cosmic blackbody background radiation (CBR) anisotropy spectrum 
\cite{Ref2}, as has been predicted on the basis of inflationary scenarios.
The best-fit models vary according to the prior assumptions made when
analyzing the data \cite{Ref3}, but together with supernova data \cite{Ref6}
suggest a model ($\Omega _{\Lambda _0}\approx 0.7,\;\Omega _{m_0}\approx
0.3,\;\Omega _0\approx 1)$ with a non-zero cosmological constant and
sufficient matter to make it almost flat, implying the universe is
cosmological-constant dominated from the present back to a redshift of about 
$z=0.326,$ and matter dominated back from then to decoupling. While the set
of models compatible with the data include those with flat spatial sections $%
(k=0)$, i.e. a critical effective energy density: $\Omega _0=1,\;$they also
include positive spatial curvature $(k=+1:\Omega _0>1\;)$ models and
negative curvature $(k=-1:\Omega _0<1)$ ones, with a weak implication that
the best-fit models have positive curvature \cite{Ref3}. This has important
implications: if true, it means that the best-fit universe models,
extrapolated unchanged beyond the visual horizon, have finite spatial
sections and contain a finite amount of matter. Whether they will expand
forever or not depends on whether the cosmological constant is
indeed constant (when these models will expand forever even though $k=+1$),
or varies with time and decays away in the far future (when they will
recollapse).

It should be noted that while inflation is taken to predict that the
universe is very close to flat at the present time, it does \emph{not} imply
that the spatial sections are \emph{exactly} flat; indeed that case is
infinitely improbable, and neither inflation nor any other known physical
process is able to specify that curvature, nor dynamically change it from
its initial value \cite{ell_ch}. Thus there is no reason to believe on the
basis of inflationary dynamics that $k=0.$ Indeed positive-curvature
universes have been claimed to have major philosophical advantages over the
flat and negatively curved cases, being introduced first by Einstein \cite
{Einstein} in an attempt to solve the problem of boundary conditions at
infinity, and then adopted as the major initial paradigm in cosmology by
Friedmann, Lemaitre, and Eddington. This view was then taken up, in 
particular by Wheeler \cite{Wheeler}, to the extent that the famous book on
gravitation he co-authored with Thorne and Misner \cite{MTW} almost
exclusively considered the positive curvature case, labeling the negatively
curved case `model universes that violate Einstein$^{\prime }$s conception
of cosmology$^{\prime }$ (see page 742). Without going that far, it is
certainly worth exploring the properties of inflationary models with $k=+1$ 
\cite{pos}, particularly as this case has been marginally indicated by some
recent observations.

In this paper we examine the dynamics of inflationary universe models (i)
with positive curvature, and (ii)\ where a cosmological constant
approximation holds in the inflationary era, deriving new limits on the
allowed numbers of e-foldings of such models as a function of the epoch when
inflation ended and of the present-day total energy density parameter $%
\Omega _0$. These limits do not contradict standard inflationary
understanding. Indeed, in a sense they enhance inflation, since 
early in the inflationary epochs of $k=+1$ universes, the deceleration 
parameter is {\it more} negative than in the $k=0$ models. We 
only model the Hot Big Bang era (post inflation to the present day) 
and the Inflationary era; our results are independent of the dynamics 
before inflation starts. Similar
results will hold for all models with only slow rolling inflation. Although
the observational evidence is that there is currently a non-zero
cosmological constant, as mentioned above, for simplicity we will consider
here only the case of an almost-flat $k=+1$ universe with vanishing
cosmological constant after the end of inflation. This approximation will
not affect the statements derived concerning dynamics up to the time of
decoupling, but will make a small difference to estimates of the number of 
e-foldings given here. We will give more accurate estimates in a more 
detailed paper on these dynamics \cite{Third}. An accompanying 
paper discusses the implications of this dynamical behaviour for 
horizons in positively-curved inflationary universes \cite{Second}.

\section{Basic Equations}

The positive-curvature Friedmann-Lema\^{i}tre (FL) cosmological model in
standard form has a scale factor $S(t)$ normalized so that the spatial
metric has unit spatial curvature at the time $t_{*}$ when $S(t_{*})=1$ (see
e.g. \cite{Wein},\cite{brazil}). The spatial sections are closed at $r-$
coordinate value increment $2\pi ;$ that is, $P=(t,r-\pi ,\theta ,\phi )$
and $P^{\prime }=(t,r+\pi ,\theta ,\phi )$ are necessarily the same point,
for arbitrary values of $t,r,\theta ,\phi ,$ and wherever the origin of
coordinates is chosen. The Hubble Parameter is $H(t)=\dot{S}(t)/S(t),$ with
present value $H_0=100h\,\,km/sec/Mpc.$ The dimensionless quantity $h$
probably lies in the range $0.7<h<0.5$.

\subsection{$k=+1$ Dynamics}

The dynamic behaviour is determined by the Friedmann equation for $k=+1$
FL universes:

\begin{equation}
\ \left( \frac{H(t)}c\right) ^2=\frac{\kappa \mu (t)+\Lambda }3-\frac
1{S(t)^2},  \label{Fried}
\end{equation}
where $\kappa $ is the gravitational constant in appropriate units and $%
\Lambda $ the cosmological constant (see e.g. \cite{Wein},\cite{brazil}).
The way this works out in practice is determined by the matter content of
the universe, whose total energy density $\mu (t)$ and pressure $p(t)$
necessarily obey the conservation equation

\begin{equation}
\dot{\mu}(t)+\left( \mu (t)+p(t)/c^2\right) 3H(t)=0.  \label{Cons}
\end{equation}
The nature of the matter is determined by the equation of state relating $%
p(t)$ and $\mu (t);$ we will describe this in terms of a parameter $\gamma
(t)$ defined by

\begin{equation}
p(t)/c^2=\left( \gamma (t)-1\right) \mu (t),\gamma \in [0,2].\ 
\label{EqnState}
\end{equation}
During major epochs of the universe's history, the matter behaviour is
well-described by this relation with $\gamma $ a constant (but with that
constant different at various distinct dynamical epochs). In particular, $%
\gamma =1$ represents pressure free matter (baryonic matter), $\gamma
=\frac 43$ represents radiation (or relativistic matter), and $\gamma =0\ $
gives an effective cosmological constant of magnitude $\Lambda =\kappa \mu $
(by equation (\ref{Cons}), $\mu $ will then be unchanging in time). In
general, $\mu $ will be a sum of such components. However we can to a good
approximation represent the universe as a series of simple epochs with only
one or at most two components in each epoch.

The dimensionless density parameter $\Omega _i(t)$ for any matter component $%
i$ is defined by 
\begin{equation}
\Omega _i(t)\equiv \frac{\kappa \mu _i(t)}3\left( \frac c{H(t)}\right) ^2\
\Rightarrow \Omega _{i0}=\frac{\kappa \mu _{i0}c^2}{3H_0^2}  \label{Omega}
\end{equation}
where $\Omega _{i0}$ represents the value of $\Omega _i(t)$ at some
arbitrary reference time $t_0$, often taken to be the present time. One can
define such a density parameter for each energy density present. We can
represent a cosmological constant in terms of an equivalent energy density $%
\kappa \mu _\Lambda =\Lambda$; from now on we omit explicit reference to $%
\Lambda ,$ assuming it will be included in this way when necessary. From
the Friedmann equation (\ref{Fried}), the scale factor $S(t)$ is related to
the total density parameter $\Omega (t),$ defined by 
\begin{equation}
\Omega (t)=\sum_i\Omega _i(t)=\frac{\kappa \mu (t)}3\left( \frac
c{H(t)}\right) ^2>1,  \label{Omega1}
\end{equation}
via the relation 
\begin{equation}
\frac 1{S(t)^2}=\left( \frac{H(t)}c\right) ^2\left( \Omega (t)-1\right)
\Rightarrow \left( \frac{H_0}c\right) ^2=\frac 1{S_0^2\left( \Omega
_0-1\right) },\;\;\Omega _0\equiv \frac{\kappa \mu _0}3\frac{c^2}{H_0^2}>1,
\label{hubble}
\end{equation}
where $\Omega _0$ is the present total value: $\Omega _0=\sum_i\Omega _{i0}$
. When $\gamma $ is constant for some component labeled $i$, the
conservation equation (\ref{Cons}) gives 
\begin{equation}
\frac{\kappa \mu _i(t)}3=\Omega _{i0}\left( \frac{H_0}c\right) ^2\left( 
\frac{S_0}{S(t)}\right) ^{3\gamma }=\frac 1{S_0^2}\left( \frac{\Omega _i{}_0 
}{\Omega _0-1}\right) \left( \frac{S_0}{S(t)}\right) ^{3\gamma }
\label{Density}
\end{equation}
for each component, on using (\ref{hubble}). The integral 
\begin{equation}
\Psi (A,B)\equiv \int_{t_A}^{t_B}\frac{cdt}{S(t)}=c\int_{S_A}^{S_B}\frac{dS}{
S\dot{S}}  \label{Psi}
\end{equation}
is the conformal time, used in the usual conformal diagrams for FL\
universes \cite{Ref4a}.

\subsection{Matter and Radiation Eras}

During the \emph{combined matter and radiation eras}, i.e. whenever we can
ignore the $\Lambda$ term in the Friedmann equation (\ref{Fried}) but
include separately conserved pressure-free matter and radiation: $\mu =\mu
_m+\mu _r$, each separately obeying (\ref{Density}), a simple analytic
expression relates $S$ and $\Psi $ \cite{brazil}. For such combined matter
and radiation, referred to an arbitrary reference point $P$ in this epoch,
we have the exact solution

\begin{equation}
S(\Psi )_P=S_P\left( \frac 12\frac{\Omega _{m_P}}{\Omega _P-1}\left( 1-\cos
\Psi \right) +\sqrt{\frac{\Omega _{r_P}}{\Omega _P-1}}\sin \Psi \right) ,
\label{HBB}
\end{equation}
where the first term is due to the matter and the second is due to
radiation. It is remarkable that they are linearly independent in this
non-linear solution. We obtain the pure radiation solution if $\Omega
_{m_p}=0,$ and the pure matter solution if $\Omega _{r_p}=0.$ The origin of
the time $\Psi $ has been chosen so that an initial singularity occurs at $%
\Psi =0,$ if the Hot Big Bang era in this model is extended as far as
possible (without an inflationary epoch). We will use this representation
from the end of inflation to the present day. It will be accurate whenever
the matter and radiation are non-interacting in the sense that their energy
densities are separately conserved, but inaccurate when they are strongly
interacting, for example when pair production takes place.

\subsection{Cosmological Constant Epoch}

During a \emph{cosmological constant-dominated era}, i.e. when $\Lambda >0$
and we can ignore matter and radiation in (\ref{Fried}), we can find the 
\emph{general solution} (with a suitably chosen origin of time) in the
simple collapsing and re-expanding form 
\begin{equation}
S(t)=S(0)\cosh \lambda t,\;\lambda \equiv c\sqrt{\frac \Lambda 3\ }
,\;S(0)\equiv \frac c\lambda\;,  \label{CC}
\end{equation}
where $t=0$ corresponds to the minimum of the radius function, i.e. the
turn-around from infinite collapse to infinite expansion, and so $S(0)$ is
the minimum value of $S(t)$ (note that we can have independent time scales
in the different eras with different zero-points, provided we match properly
between eras as discussed next). This is of course just the de Sitter
universe represented as a Robertson-Walker spacetime with positively-curved
space sections \cite{Schr} , and can be used to represent an inflationary
universe era for models with $k=+1\;$if we restrict ourselves to the
expanding epoch $\;$ 
\begin{equation}
t\geq t_i\geq 0  \label{t0}
\end{equation}
for some suitable initial time $t_i\ $which occurs after the end of the
Planck era, so $t_i\geq t_{Planck}$. We will represent the inflationary era
(preceding the Hot Big Bang era) in this way.

The Hubble parameter is $H(t)=\lambda \tanh \lambda t$ (zero at $t=0$ and
positive for $t>0$) and the density parameter $\Omega _\Lambda (t)$ is given
by 
\begin{equation}
\;\Omega _\Lambda (t)=\frac \Lambda 3\frac{c^2}{H^2(t)}=\frac 1{(\tanh
\lambda t)^2},  \label{Consts1}
\end{equation}
which diverges as $t\rightarrow 0$ and tends to $1$ as $t\rightarrow \infty
.\ $The inflationary effect is enhanced in such models as compared with $k=0$
models, because the deceleration parameter $q=-\ddot{S}/(SH^2)$ is here even
more negative than in those scale-free models.

\subsection{Joining Different Eras}

Junction conditions required in joining two eras with different equations of
state are that we must have $S(t)$ and $\dot{S}(t)$ continuous there, thus $%
H(t)$\ is continuous also. By the Friedmann equation this implies in turn
that $\mu (t)$ is continuous, so by its definition $\Omega (t)$ is also
continuous (note that it is $p(t)$ that is discontinuous on spacelike
surfaces of discontinuity). We need to demand, then, that any two of these
quantities are continuous where the equation of state is discontinuous; for
our purposes it will be convenient to take them as $S(t)$ and $\Omega (t).$
Thus we need to know $S(t)$ and $\Omega (t)$ at the beginning and end of
each era to get a matching with $S_{-}=S_{+}$ and $\Omega _{-}=\Omega _{+}$.

The matching we need to perform is between the Hot Big Bang era and the
Inflationary era. Now for combined matter and radiation, referred to an
arbitrary reference point $P$, we have (\ref{HBB}). Writing the same
solution in the same form (with the initial singularity at $\Psi =0$ in both
cases) but referred to another reference point $Q,$ we have the same
expressions but with $P$\ replaced by $Q$ everywhere. As these are the same
evolutions referred to different events, $S(\Psi )_P=S(\Psi )_Q$ for all $%
\Psi ,$ so they must have identical functional forms. Matching the two
expressions for all $\Psi ,$ the coefficients for $\cos \Psi $ and $\sin
\Psi $ on each side must separately be equal. Letting $S_P/S_Q=\mathcal{R},$
this gives the total density parameter $\Omega _Q(\mathcal{R})=\Omega _{mQ}( 
\mathcal{R})+\Omega _{rQ}(\mathcal{R})$ at the event $Q$ in terms of the
density parameter values at $P.$ Taking the event $Q$ to be the end of
inflation and the event $P$ to be here and now given by $t=t_0$, we find 
\begin{equation}
\Omega _Q(\mathcal{R})=\frac{\,\mathcal{R}\,\left( \Omega _{m0}+\,\mathcal{R}
\Omega _{r0}\right) }{\,\mathcal{R}^2\Omega _{r0}+\,\mathcal{R}\,\Omega
_{m0}-(\Omega _{m0}+\Omega _{r0}-1)}.  \label{match}
\end{equation}
The matching condition is then given by assuming $\Omega _\Lambda
(t_Q)=\Omega _Q(\mathcal{R})$. We are assuming here that the details of
reheating at the end of inflation are irrelevant: conservation of total
energy must result in the total value of $\Omega $ at $Q$ being constant
during any such change (one can easily modify this condition if desired).
This gives $\Omega _\Lambda (t_Q)$ at the end of inflation $Q$ by 
(\ref{Consts1}), which then gives $t_Q$ in the inflationary epoch 
described by (\ref{CC}): 
\begin{equation}
\;t_Q=\frac 1\lambda \mathrm{arctanh}\frac 1{\sqrt{\Omega _\Lambda (t_Q)}
}=\frac 1\lambda \mathrm{arctanh}\frac 1{\sqrt{\Omega _Q(\mathcal{R})}}\;,
\label{Consts2}
\end{equation}
with $\Omega _Q(\mathcal{R})$ given by (\ref{match}). Note that in these
expressions, the ratio $\mathcal{R=}S_0/S_Q$ is the expansion ratio from the
end of inflation until today.

\section{Parameter Limits}
\subsection{Maximum Number of e-foldings:\ $N_{\max }$}

The maximum number of e-foldings available until time $t_Q$ with the given
value $\Omega _\Lambda (t_Q)$ is given by the expansion form (\ref{CC}) with 
$t=t_Q$ given by (\ref{Consts2}): 
\begin{equation}
e^{N_{\max }}=\frac{S(t_Q)}{S(0)}=\cosh \left( \mathrm{arctanh}\frac 1{\sqrt{
\Omega _Q(\mathcal{R})}}\right) ,  \label{Nfold}
\end{equation}
because $S(0)$ is the minimum value of $S(t)$ (by our choice of time
coordinate for this era, the throat where expansion starts is set at $t=0$)
and $t$ is restricted by (\ref{t0}). If the universe starts off at any time
later than $t=0$ in the expanding era $t>0$, there will be fewer e-foldings
before the end of inflation $t_Q$.

Defining the difference of $\Omega _0$ from unity to be $\delta >0$: 
\begin{equation}
\Omega _0=1+\delta \Leftrightarrow \Omega _{m0}=1+\delta -\Omega _{r0}\;,
\label{delta}
\end{equation}
we find from (\ref{Nfold}) and (\ref{match}) that the maximum number of
inflationary e-foldings that can occur before inflation ends at an event $Q$
with expansion ration $\mathcal{R}$, is 
\begin{equation}
N_{\max }(\mathcal{R},\delta )=\ln \left[ \cosh \left( \mathrm{arctanh}\frac{%
\sqrt{\alpha -\delta }}{\sqrt{\alpha }}\right) \right] =\frac 12\ln \alpha
-\frac 12\ln \delta\;,  \label{N-Limit1}
\end{equation}
where 
\[
\alpha \equiv \mathcal{R}^2\Omega _{r0}+\,\mathcal{R(}\,1+\delta -\Omega
_{r0}). 
\]
This e-folding limit essentially represents a matching of the present day
radiation density $\Omega _{r0}$ to the energy density limits that may be
imposed at the end of inflation, which will place restrictions on the
possible value of the expansion ratio $\mathcal{R}$. It does not take into
account matter-radiation conversions in the Hot Big Bang era, which we
consider in a later section.

\subsection{Density parameter variation from unity: $\delta $}

What are the implications? The inversion of (\ref{N-Limit1}) in terms of $%
\delta $ is 
\begin{equation}
\delta (\mathcal{R},N_{\max })=\mathcal{R}\frac{\mathcal{R}\Omega
_{r0}+1-\Omega _{r0}}{e^{2N_{\max }}-\mathcal{R}}.  \label{delta_limit}
\end{equation}
Note that this value diverges when $N_{\max }=\frac 12\ln (\mathcal{R})$, so
this is the value corresponding to a turn-around today ($\Omega _0=\infty
\Leftrightarrow H_0=0$). Thus there is a smallest value for $N_{\max }$ for
each set of parameters $\Omega _{r0},\mathcal{R}.$

\subsection{Expansion ratio since end of inflation: $\mathcal{R}$}

The inversion in terms of the ratio $\mathcal{R}$ is

\begin{equation}
\mathcal{R(}\delta ,N_{\max })=\frac 1{2\Omega _{r0}}\left( \sqrt{4\Omega
_{r0}\delta e^{2N_{\max }}+\mathcal{(}\,1+\delta -\Omega _{r0})^2}-\mathcal{%
\ ( }\,1+\delta -\Omega _{r0})\right) .  \label{R_Limit}
\end{equation}
For large $\delta e^{2N_{\max }}$ this is well-approximated by 
\begin{equation}
\mathcal{R(}\delta ,N_{\max })=\frac 1{\sqrt{\Omega _{r0}}}\sqrt{\delta }
e^{N_{\max }}.  \label{R_large}
\end{equation}

\subsection{Numerical Values}

The epoch chosen for the end of inflation will determine the expansion
parameter $\mathcal{R}$. What is a realistic expectation for the end of
inflation? A typical figure for the energy then is $10^{14}$Gev, just below
the GUT energy. In terms of temperature this is equivalent to $%
T=1.16\times 10^{27}K$ at the end of inflation. 
But the CBR temperature is $2.75K$ today, so assuming that in the 
Hot Big Bang era $T$ scales as $1/S(t)$, we
obtain the value $\mathcal{R=}1.16\times 10^{27}/2.75=4.22\times 10^{26}.$ This
can be taken as an upper value (inflation ends below the GUT energy), but
requires correction for pair production processes at high temperatures (see
below). An absolute lower value would be $\mathcal{R=}10^{12}$ (ensuring
that inflation ends before baryosynthesis and nucleosynthesis begin).
Finally, how many e-foldings would be expected during inflation? A value
demanded in most inflationary scenarios is at least $N=60$, required firstly
to smooth out the universe, and then assumed in the usual structure
formation studies. A typical figure is an expansion ratio $e^N=$ $%
e^{70}=2.5\times 10^{30}$ (\cite{Pad}, p.355); some studies quote much
higher values for $N$. The value of the difference from flatness $\delta $
today (\ref{delta}) is probably in the range $-0.05<\delta <0.1$; it might
be very small indeed, as assumed in many inflationary scenarios. In this
paper, we are only studying the case $\delta >0$ because we are assuming
positive spatial curvature. The Cosmic Background Radiation density today is
well-determined from its temperature of $T=2.75K$, and is $\Omega
_{r0}=4.2\times 10^{-5}h^{-2}$ \cite{Pad}, because we include the neutrino degrees
of freedom here. Taking $h=0.65,$ this gives the value $\Omega _{r0}\simeq
10^{-4}.$

We now explore the effect of variation of all these parameters except $%
\Omega _{r0},$ which we take as fixed, because the temperature of that
radiation is extremely well determined by observation. It is this quantity
that then determines the numbers in what follows (if we did not fix this
number, we would get only functional relations but not specific numerical
limits on what can happen). There will be a small variation in the results
that follow if we vary $h,$ because the CBR temperature is converted into
an equivalent $\Omega _{r0}$ value by the present value of the Hubble
constant (expressed in terms of $h$).

\subsection{Allowed end of inflation}

In terms of the ratio $\mathcal{R},$ for $\Omega _0=1+\delta $ and using the
above value for $\Omega _{r0}$, we can get at most $N$  e-foldings during
inflation if the end of inflation occurs at the expansion ratio 
\begin{equation}
\mathcal{R(}\delta ,N_{\max })=\frac 1{8.4*10^{-5}h^{-2}}\left( \sqrt{
16.8*10^{-5}h^{-2}\delta e^{2N_{\max }}+\Delta ^2}-\Delta \right) ,\;
\end{equation}
\[
\Delta \equiv (\,1+\delta -4.2\times 10^{-5}h^{-2})\;. 
\]
For $N_{\max }\geq 60$ and $\delta >10^{-4}$ we can use the simple
approximation 
\begin{equation}
\mathcal{R}(\delta ,N_{\max })=154.3h\sqrt{\delta }e^{N_{\max }}.
\label{R_large1}
\end{equation}
Values for $N_{\max }=60$ and $h=0.65$ are
\vspace{3ex}\\
\begin{center}
\begin{tabular}{|c|c|c|c|c|} \hline
$\delta $ & $0.0001$ & $0.0005$ & $0.0008$ & $0.001$ \\ \hline
$\mathcal{R}$ & $1.45\times 10^{26}$ & $2.56\times 10^{26}$ & $3.24\times
10^{26}$ & $3.62\times 10^{26}$ \\ \hline
$\delta $ & $0.01$ & $0.1$ & $1$ & $2$ \\ \hline
$\mathcal{R}$ & $1.15\times 10^{27}$ & $3.62\times 10^{27}$ & $1.46\times
10^{28}$ & $1.62\times 10^{28}$ \\ \hline
$\delta $ & $3$ & $4$ & $5$ & $10$ \\ \hline
$\mathcal{R}$ & $1.98\times 10^{28}$ & $2.29\times 10^{28}$ & $2.56\times
10^{28}$ & $3.62\times 10^{28}$ \\ \hline
\end{tabular}
\vspace{3ex}\\
\end{center}
Now as commented above, we do not want to exceed the value $\mathcal{R}
=4.22\times 10^{26}$ corresponding to the GUT energy density. The conclusion
is that we exceed this value if $\delta >0.005.$ The limit will become
stronger if we demand more e-foldings.

\subsection{Allowed density range today}

Assume now $\mathcal{R=}4.22\times 10^{26}$ as in standard 
texts \cite{Pad}. Then, neglecting a small term we obtain
\[
\delta (N_{\max })=\frac{7.47\times 10^{48}}{h^2\left( e^{2N_{\max
}}-4.22\times 10^{26}\right) }\;. 
\]
In this case, the smallest number $N_{\max }$ is given by $e^{2N_{\max
}}=4.22\times 10^{26}$, and so $N_{\max }>30.65$. For various inflationary
e-folding values $N_{\max }$ greater than this amount, we find, on setting $%
h=0.65$:
\vspace{3ex}\\
\begin{center}
\begin{tabular}{|c|c|c|c|c|} \hline
$N_{\max }$ & $40$ & $50$ & $55$ & $56$ \\ \hline
$\delta $ & $3.19\times 10^{14}$ & $6.58\times 10^5$ & $29.86$ 
& $4.04$ \\ \hline
$N_{\max }$ & $57$ & $58$ & $59$ & $60$ \\ \hline
$\delta $ & $0.55$ & $7.40\times 10^{-2}$ & $1.00\times 10^{-2}$ & 
$1.36\times 10^{-3}$ \\ \hline
\end{tabular}
\vspace{3ex}\\
\end{center}
We see here the very sharp decline as $N_{\max }$ increases through $56$ to $%
59$. Values higher than $58$ strongly limit the value of $\delta $ , i.e.
the allowed domain in the ($\Omega _\Lambda ,\Omega _M)$ plane.

\subsection{Allowed e-foldings}

Finally again assuming $\mathcal{R=}4.22\times 10^{26}$ , the maximal number
of e-foldings is given by 
\begin{equation}
N_{\max }(\delta )=30.654+\frac 12\ln \left( 1.7724\times 10^{22}+h^2+\delta
h^2\right) -\allowbreak \ln h-\frac 12\ln \delta\;.
\end{equation}

So for various values of $\delta ,$ if $h=0.65$, we find the allowed number
of e-foldings: 
\vspace{3ex}\\
\begin{center}
\begin{tabular}{|c|c|c|c|} \hline
$\delta $ & $0.00001$ & $0.0001$ & $0.001$ \\ \hline 
$N_{max}$ & $62.46$ & $61.31$ & $60.15$ \\ \hline
$\delta $ & $0.01$ & $0.1$ & $1$ \\ \hline
$N_{\max }$ & $59.00$ & $57.85$ & $56.70$ \\ \hline 
$\delta $ & $2$ & $4$ & $10$ \\ \hline
$N_{\max }$ & $56.35$ & $56.01$ & $55.55$ \\ \hline
\end{tabular}
\vspace{3ex}\\
\end{center}
So we again see the here the crucial e-folding range $56$ to $58$ as the
limit allowing substantial values of $\delta .$ This range is less than that
normally assumed for the end of inflation.

\section{Actual Number of e-foldings:\ $N$}

The actual number of e-foldings during the inflationary era until time $t_Q,$
with the $\Omega $-value $\Omega _\Lambda (t_Q),$ starting from time $t_i,$
with the $\Omega -$value $\Omega _\Lambda (t_i),$ is 
\begin{equation}
e^N=\frac{S(t_Q)}{S(t_i)}=\frac{\cosh \left( \mathrm{arctanh}\frac 1{\sqrt{
\Omega _Q}}\right) }{\cosh \left( \mathrm{arctanh}\frac 1{\sqrt{\Omega _i}
}\right) }=\sqrt{\frac{\Omega _i-1}{\Omega _i}}\sqrt{\frac{\Omega _Q}{\Omega
_Q-1}},  \label{Nfold1}
\end{equation}
( from (\ref{CC},\ref{Consts1})), that is 
\begin{equation}
N(\Omega _Q,\Omega _i)=\frac 12\ln \left\{ \frac{\Omega _Q\left( \Omega
_i-1\right) }{\Omega _i\,(\Omega _Q-1)}\right\} .  \label{Nfold2}
\end{equation}
The solution for $\Omega _Q$ is 
\[
\Omega _Q(N,\Omega _i)=\frac{e^{2N}\Omega _i}{e^{2N}\Omega _i-\Omega _i+1}, 
\]
so using (\ref{match}), substituting $\Omega _0=1+\delta $ , and solving for 
$\delta $ gives: 
\begin{equation}
\delta (\mathcal{R},N,\Omega _i)=\frac{\mathcal{R}\left\{ \mathcal{R}\Omega
_{r0}+1-\Omega _{r0}\right\} (\Omega _i-1)}{e^{2N}\Omega _i-\,\mathcal{R}
(\Omega _i-1)}.  \label{delta2}
\end{equation}
This gives the standard result that inflation through $N$ e-foldings
decreases $\delta ,$ and can indeed make it arbitrarily small if $N$ is
large enough. The limit giving $N_{\max }$ is the irregular limit: $\Omega
_i\longrightarrow \infty .$ We obtain a minimum allowed number of e-foldings
from the requirement that $e^{2N}\Omega _i>\,\mathcal{R}(\Omega _i-1).$ This
gives $N_{\min }$ (the value when $\delta $ diverges) to be

\begin{equation}
N_{\min }(\mathcal{R},\Omega _i)=\frac 12\ln \left( \frac{\mathcal{R(}\Omega
_i-1)}{\Omega _i}\right) .  \label{Nmin}
\end{equation}
Universes with less e-foldings will have collapsed before today.

\section{Implications}
\%

We have arrived at the following interesting result: Consider a universe 
with a \emph{cosmological constant dominated inflationary epoch, }
where inflation ends by $10^{14}GeV$. Then, noting that $\Omega
_0>1\Rightarrow k=+1,$ we find that with our assumptions above, if $\Omega
_0>1.01,$ the limits above apply in our multi-stage simple model and there 
cannot have been inflation through $60$ e-foldings or more in such a model. 
Thus for example $\Omega _0=1.01$ contradicts the possibility of an 
exponentially expanding inflationary scenario with more than 60 e-foldings 
in our past in such a model. This is essentially because the curvature
enhances the effect of inflation in the very early universe, making the
curve $S(t)$ bend up more than it would have done in the zero-curvature case
and resulting in $\Omega $ diverging at a turn-around point if the
inflationary era is extended too far to the past. This is disallowed by the
instability of a collapsing inflationary epoch \cite{collapse}. 

However these values depend on the assumptions we make for $\mathcal{R}$ 
and $h$ in this simple multi-stage model, and would be changed by more 
accurate models; there will be variations of these figures with detailed 
inflationary scenarios and more accurate modeling. In particular, we have 
carried out preliminary estimates of the effects of (a) changing 
matter-radiation relations in the hot big bang era, due to pair 
creation and extra degrees of freedom arising; these seem to make little 
difference; and (b) the effect of a previous radiation dominated era at 
the start of inflation, resulting in an initial inflationary era where 
radiation was non-negligible. The basic effect would remain in this case, 
but the numbers estimated above would change. These refinements will be 
considered in a paper \cite{Third} examining the relevant dynamics in more 
detail.
 
The main point of this paper is that such limits exist and should be 
taken into account when examining inflationary models with $k=+1$. 
The limits given above are only for the simple model 
considered here; they will be different in more detailed models.

\subsection{Criterion for this to happen}

This calculation is for an epoch of inflation driven by a cosmological
constant. However there are numerous other forms of inflation. The key point
then is that similar effects will occur in all inflationary models in which
the effective energy density of the scalar field varies more slowly than the
curvature term in the Friedmann equation, which varies as $S^{-2}.$ 
From (\ref{Density}), this will happen if $3\gamma <2.$ The limiting behaviour
where the energy density mimics the curvature term is a coasting universe
with $3\gamma =2\Leftrightarrow $ $\mu +3p/c^2=0$. Scalar fields can give
any effective $\gamma $ from $0$ to $2$, so there will be fast-rolling
scalar-field driven models where $3\gamma >2$ . However these will not then
be inflationary, for they will not be accelerating (the requirement for an
accelerating universe is $\mu +3p/c^2<0)$. Thus \emph{effects of the kind
considered here will occur in all positive curvature inflationary universes}, 
but power-law models will have different detailed behaviour than the ones with 
an effective cosmological constant calculated above. The numbers 
will be different and the constraints may be much less severe.

\section{Conclusion}

If we ever \textit{observationally determine} that $\Omega _0>1\Rightarrow $ 
$k=+1,$ then $\delta >\delta _0$ where $\delta _0$ is some value
sufficiently large that we can distinguish the value of $\Omega _0$ from
unity, and so will certainly be greater than $0.01$ (for otherwise we 
could not observationally prove that $\Omega_0>1$). Thus there cannot in
this case have been exponential inflation through some value that will 
depend on the model used; in the case considered above, it is about $59$
e-foldings, so such an inflationary scenario, with $60$ or more e-foldings,
could not have occurred. Hence it is of considerable interest to try all
forms of cosmological tests to determine if $\Omega _0>1.$ It is of course
possible we will never determine observationally whether $\Omega _0>1$ or $%
\Omega _0<1.$ The point of this paper is to comment that there are
substantial dynamical implications if we can ever make this distinction on
the basis of observational data. There is not a corresponding implication on
the negative side, i.e. for $\delta <0\Leftrightarrow \Omega _0<1$ (one
might then claim limits on the number of e-foldings caused by limits on $%
\Omega _{Planck}$ or $H_{Planck}$ at the end of the Planck time; but the
results presented here are independent of any such considerations)$.$ Thus
if we could ever determine say that $\Omega _0=0.99,$ this would not imply
any limit on the number of e-foldings, whereas for $\Omega _0=1.01,$ such
restrictions are implied.

Many inflationary theorists would not find this conclusion surprising, as
they would expect the final value of $\delta$ to be very small, as is
indicated here, and would assume that if we were too far from flat today
this was just because, given the starting conditions for the inflationary
era, one had not had enough e-foldings to truly flatten the universe; so
more e-foldings should be employed, and we would end up much closer to flat
today. However they have arrived at that conclusion by examining the case of
scale-free (exponential) expansion, which arises when the spatial curvature
term in the Friedmann equation is ignored, and then placing bounds on the
value of the allowed energy density at the start of inflation. But the point
of the present analysis is precisely that one cannot ignore that curvature 
term at early enough times in an inflationary epoch driven by a cosmological
constant. It is the resulting non-scale-free behaviour that leads to the
restrictions on allowed e-foldings calculated above, irrespective of the
initial conditions inherited from the Planck era. The implication is that if
you call up the extra e-foldings needed for that programme just outlined,
and end up consistent with the presently observed CBR\ density, then
necessarily a limit such as $\Omega _0<1.001$ holds. Thus this kind of 
result strengthens the inflationary intuition. However that e-folding limit 
is not incorporated in the models usually used to calculate the CBR anisotropy.

Indeed if $\Omega _0>1$, so that only restricted e-foldings can occur and be
compatible with the observed CBR temperature, this could have significant
effects on structure formation scenarios. The usual analyses resulting in
the famous observational planes with axes $\Omega _m$ and $\Omega _\Lambda$ 
\cite{Ref3} are based on assuming that more than $60$ e-foldings can occur
even if $k=+1$. We suggest the theoretical results need re-examination in
the domain where $k=+1$ and only a restricted number of e-foldings can
occur. The major point is that the dynamical behaviour is discontinuous in
that plane: as $\Omega _0$ varies from $1-\delta $ to $1+\delta $, however
small $\delta $ is, the curvature sign $k$ changes from $-1$ to $+1$ and the
corresponding term $k/S(t)^2$ in the Friedmann equation - which necessarily
dominates over any constant term in that equation, for small $S(t)$ -
completely changes in its effects. When $k=+1$ it eventually causes a
turn-around for some $t_0$; when $k=-1$ it hastens the onset of the initial
singularity.

It should be noted that this conclusion is based purely on examining
inflation in FL\ universe models with a constant vacuum energy, and is not
based on examinations of pre-inflationary or Trans-Planckian physics on the
one hand, nor on studies of embedding such a FL\ region in a larger region
on the other. It is
based solely on the dynamics during the inflationary epoch. However it
considers only a constant vacuum energy, equivalent to a no-rolling
situation, and so does not take scalar field dynamics properly into account.
It will be worth examining slow-rolling and fast-rolling models to see what
the bounds of behaviour for $k=+1$ inflationary models are in those cases.
We indicated above that insofar as these universes are inflationary (i.e.
they are accelerating during the scalar-field dominated era), similar
e-folding bounds may be expected in these cases also. Also as indicated above, the results will be modified if there is a substantial radiation
density during the initial phase of inflation. We are currently investigating the difference that this will make.

We are fully aware that in order to properly study the issue, we need to
examine anisotropic and inhomogeneous geometries rather than just FL\
models, because analyses based on FL\ models with their Robertson-Walker
geometry cannot be used to analyse very anisotropic or inhomogeneous eras.
Nevertheless this study shows there are major dynamical differences in
inflationary FL universes with $k=+1$ or $k=0$. The implication is (a) that
we need to try all observational methods available to determine if $k=+1$,
because this makes a significant difference not only to the spatial
topology, but also to the dynamical and causal structure of the universe,
and (b) we should examine inhomogeneous inflationary cosmological models to
see if any similar difference exists between models that are necessarily
spatially compact, and the rest.

We thank Roy Maartens, Bruce Bassett, and Claess Uggla for useful comments, 
and the NRF (South Africa) for financial support.

\end{document}